\documentstyle[aps,twocolumn,epsfig]{revtex}
\begin{document}
\title{  $J/\psi$  suppression in Pb+Pb collisions and $p_T$ broadening}
\author{\bf A. K. Chaudhuri\cite{byline}}
\address{ Variable Energy Cyclotron Centre\\
1/AF,Bidhan Nagar, Calcutta - 700 064\\}
\maketitle
\begin{abstract}

We  have  analysed the NA50 data, on the centrality dependence of
$p_T$ broadening of  $J/\psi$'s,  in  Pb+Pb  collisions,  at  the
CERN-SPS.  The  data  were  analysed  in a QCD based model, where
$J/\psi$'s are suppressed in 'nuclear' medium. Without  any  free
parameter,  the  model  could  explain  the NA50 $p_T$ broadening
data. The data were also analysed in a QGP based threshold model,
where $J/\psi$ suppression is 100\% above a critical density. The
QGP based model could not explain the NA50 $p_T$ broadening data.
We have also predicted  the  centrality  dependence  of  $J/\psi$
suppression and $p_T$ broadening at RHIC energy. 
Both the models, the QGP based threshold model and the QCD based
nuclear absorption model, predict $p_T$ broadening very close to
each other. 

\end{abstract}


Since  the prediction by Matsui and Satz \cite{ma86} that binding
of a $c\bar{c}$ pair into a $J/\psi$ meson will  be  hindered  in
quark-gluon  plasma  (QGP), $J/\psi$ suppression is recognized as
an important tool for the identification of  the  possible  phase
transition from confined to deconfined matter. NA50 collaboration
measured  centrality  dependence of $J/\psi$ suppression in 158 A
GeV Pb+Pb collisions \cite{na50a}. They observed suppression well
beyond the standard nuclear absorption model. Initially the  data
were  interpreted  in  terms  of successive melting of charmonium
states in QGP \cite{na50a}. However, later, it was realized  that
the  data  could  be  explained  in  a  variety of models with or
without  QGP  \cite{bl00,ch01,ca00,ch02,ch02a}.  What   it   more
intriguing   is  that  the  predicted  centrality  dependence  of
$J/\psi$ at RHIC energy in a model without QGP, matches with  the
model  prediction  with  QGP  \cite{ch02}.  It  appears  that the
$J/\psi$ suppression may not be a good signal for the deconfining
phase transition.

Apart  from  the  centrality  dependence of $J/\psi$ suppression,
NA50 collaboration also presented data on the  transverse  energy
(centrality)   dependence   of  $p_T$  broadening  of  $J/\psi$'s
\cite{na50b}, which did not receive much attention.  Kharzeev  et
al  \cite{kh97}  suggested  that $p_T$ broadening of $J/\psi$ can
distinguish between QGP and nuclear matter. They argued that in a
nuclear matter, $p_T$ broadening will saturate at large $E_T$. In
contrast, in a QGP, $p_T$ broadening will visibly decrease.  They
used  conventional  Glauber model of nuclear absorption, which we
know,  could  not  explain  the  NA50  data  on  the   centrality
dependence  of  $J/\psi$  over  Drell-Yan  ratio.  The centrality
dependence of $J/\psi$  suppression  is  well  explained  in  the
'unconventional'  QCD based nuclear absorption model \cite{ch02}.
In the present letter we have analysed the NA50 $p_T$  broadening
data  in  the  model. Without any free parameter, the model could
explain the data. For comparison purpose, we  have  analysed  the
data also in the QGP based threshold model \cite{bl00}. The model
fails to fit the data. We have also predicted $p_T$ broadening at
RHIC  energy. Interestingly, both the models predict very similar
$p_T$ broadening at  RHIC.  It  seems  that,  like  the  $J/\psi$
suppression,   centrality   dependence  of  $p_T$  broadening  of
$J/\psi$ may not be a  good  signal  for  the  deconfining  phase
transition.

It  is  well  known  that  in pA and AA collisions, the secondary
hadrons generally shows a $p_T$ broadening. The natural basis for
this broadening is the  initial  state  parton  scatterings.  For
$J/\psi$'s,   gluon  fusion  being  the  dominant  mechanism  for
$c\bar{c}$  production,   initial   state   scattering   of   the
projectile/target  gluons  with  the  target/projectile  nucleons
causes the intrinsic momentum broadening of the gluons, which  is
reflected  in the $p_T$ distribution of the resulting $J/\psi$'s.
Parametrizing the  intrinsic  transverse  momentum  of  a  gluon,
inside a nucleon as,

\begin{equation} f(q_T) \sim exp(-q^2_T/<q^2_T>) \end{equation}

\noindent  momentum  distribution of the resulting $J/\psi$ in NN
collision is obtained by convoluting two such distributions,

\begin{equation}             f^{J/\psi}_{NN}(p_T)            \sim
exp(-p^2_T/<p^2_T>_{NN}) \end{equation}

\noindent    where    $<p^2_T>_{NN}    =   <q^2_T>+<q^2_T>$.   In
nucleus-nucleus collisions at  impact  parameter  ${\bf  b}$,  if
before  fusion, a gluon undergo random walk and suffer $N$ number
of subcollisions, its square momentum  will  increase  to  $q^2_T
\rightarrow  q^2_T  +  N\delta_0$,  $\delta_0$  being the average
broadening in each subcollisions.  Square  momentum  of  $J/\psi$
then easily obtained as,

\begin{equation}     \label{1}     <p^2_T>^{J/\psi}_{AB}(b)     =
<p^2_T>_{NN} + \delta_0 N_{AB}({\bf b}) \end{equation}

\noindent  where $N_{AB}({\bf b})$ is the number of subcollisions
suffered by the projectile and target gluons with the target  and
projectile nucleons respectively.

Average number of collisions $N_{AB}({\bf b})$ can be obtained in
a  Glauber  model.  At  impact parameter ${\bf b}$, the positions
$({\bf s},z)$ and $({\bf b-s},z^\prime)$ specifies the  formation
point  of  $c\bar{c}$  in  the  two nuclei, with ${\bf s}$ in the
transverse plane and $z,z^\prime$ along the beam axis. The number
of collisions, prior to $c\bar{c}$ pair formation, can be written
as,

\begin{eqnarray}  \label{2}  N(b,s,z,z^\prime)  =  && \sigma_{gN}
\int_{-\infty}^z dz_A \rho_A(s,z_A) \\ \nonumber && + \sigma_{gN}
\int_{-\infty}^{z^\prime}        dz_B        \rho_B(b-s,z^\prime)
\end{eqnarray}

\noindent where $\sigma_{gN}$ is the gluon-nucleon cross section.
Above  expression  should  be  averaged  over  all  positions  of
$c\bar{c}$ formation with  a  weight  given  by  the  product  of
nuclear densities and survival probabilities $S$,

\begin{eqnarray}\label{3}     &&N_{AB}(b)=     \int     d^2     s
\int^\infty_{-\infty}   dz   \rho_A(s,z)    \int^\infty_{-\infty}
dz^\prime    \rho_B(b-s,z^\prime)    \times   \\   \nonumber   &&
S(b,s,z,z^\prime)     N(b,s,z,z^\prime)     /      \int      d^2s
\int^\infty_{-\infty}  dz  \rho_A(s,z)  \times  \\  \nonumber  &&
\int^\infty_{-\infty}       dz^\prime        \rho_B(b-s,z^\prime)
S(b,s,z,z^\prime) \end{eqnarray}

Finally,  corresponding quantity at fixed transverse energy $E_T$
is obtained as,

\begin{eqnarray}\label{4a}  N_{AB}(E_T)  =  &&\int d^2 b P(b,E_T)
\sigma_{AB} N_{AB}(b)  /  \\  \nonumber  &&  \int  d^2b  P(b,E_T)
\sigma_{AB} \end{eqnarray}

\noindent  where $\sigma_{AB}$ is the inelastic cross section for
AB collisions. $P(b,E_T)$ is the $E_T-b$ correlation function. We
have used the Gaussian form for the $E_T-b$ correlation,

\begin{equation}\label{5}             P(b,E_T)            \propto
exp(-(E_T-qN_p(b))^2/2q^2aN_p(b)) \end{equation}

\noindent where $N_p(b)$ is the number of participant nucleons at
impact  parameter  b.  $a$  and  $q$  are  parameters  related to
dispersion and average transverse energy.  For  Pb+Pb  collisions
the parameters are, $a$=1.27 and $q$=0.274 GeV \cite{bl00}.

Survival  probability  $S(b,s,z,z^\prime)$ in Eq.\ref{3} is model
dependent.  We  have  calculated   it   using   our   QCD   based
'unconventional'   nuclear  absorption  model  \cite{ch02,qiu98}.
Briefly, $J/\psi$ production is assumed to be a two step process,
(a)  formation  of  a  $c\bar{c}$  pair,  which   is   accurately
calculable  in QCD and (b) formation of a $J/\psi$ meson from the
$c\bar{c}$  pair,  which  is  conveniently   parameterized.   The
$J/\psi$  cross  section  in  $AB$  collisions, at center of mass
energy $\sqrt{s}$ was then written as,

\begin{eqnarray} \sigma^{J/\psi} (s) &&
=K \sum_{a,b} \int dq^2 \left( \frac{\hat \sigma_{ab \rightarrow
cc}} {Q^2} \right) \int dx_F \phi_{a/A}(x_a,Q^2) \\ \nonumber
&&   \phi_{b/B}(x_b,Q^2)   \frac{x_a   x_b}{x_a   +  x_b}  \times
F_{c\bar{c} \rightarrow J/\psi} (q^2), \end{eqnarray}

\noindent  where  $\sum_{a,b}$  runs over all parton flavors, and
$Q^2 = q^2 +4 m_c^2$. The  $K$  factor  takes  into  account  the
higher  order corrections. The incoming parton momentum fractions
are fixed by kinematics and are $x_a
=(\sqrt{x^2_F+4Q^2/s}+x_F)/2$               and              $x_b
=(\sqrt{x^2_F+4Q^2/s}-x_F)/2$.
$\hat  \sigma_{ab \rightarrow c\bar{c}}$ are the subprocess cross
section and are given in \cite{be94}. $F_{c  \bar{c}  \rightarrow
J/\psi}(q^2)$  is  the  transition  probability that a $c\bar{c}$
pair with relative momentum square $q^2$ evolve into  a  physical
$J/\psi$ meson. It is parameterized as,

\begin{eqnarray} \label{4} F_{c \bar{c} \rightarrow J/\psi} (q^2)
= && N_{J/\psi} \theta(q^2) \theta({4m^\prime}^2 - 4 m_c^2 -q^2) \\ \nonumber
&&   (1  -  \frac{q^2}{{4m^\prime}^2  -  4  m_c^2  })^{\alpha_F}.
\end{eqnarray}

In  a  nucleon-nucleus/nucleus-nucleus  collision,  the  produced
$c\bar{c}$ pairs interact with nuclear medium before  they  exit.
It  is  argued  that  the  interaction  of a $c\bar{c}$ pair with
nuclear environment increases the square of the relative momentum
between the $c\bar{c}$ pair. As a result, some of the  $c\bar{c}$
pairs  can  gain  enough  relative  square  momentum to cross the
threshold to become an open charm meson. Consequently, the  cross
section  for  $J/\psi$  production  is reduced in comparison with
nucleon-nucleon cross section. If the  $J/\psi$  meson  travel  a
distance  $L$, $q^2$ in the transition probability is replaced to
$q^2 \rightarrow q^2 +\varepsilon^2 L$, $\varepsilon^2$ being the
relative square momentum gain per unit length. Parameters of  the
model  were  fixed from experimental data on total $J/\psi$ cross
section in pA/AA collisions. It is thus essentially  a  parameter
free calculation for Pb+Pb collisions.

Fluctuations  of  $E_T$  at  a  fixed  impact  parameter plays an
important role in $J/\psi$ suppression in Pb+Pb  collisions.  The
2nd  drop  in the $J/\psi$ over Drell-Yan ratio at 100 GeV is due
these fluctuations only. Fluctuations of $E_T$ at a fixed  impact
parameter   also   affect   the   average  number  of  collisions
$N_{AB}(E_T)$. As will be shown later, it plays an important role
in explaining the NA50 $p_T$ broadening data. We have taken  into
account the $E_T$ fluctuations by the replacement,

\begin{equation}  N_{AB}(b)  \rightarrow  E_T/<E_T>(b) N_{AB}(b).
\end{equation}

\begin{figure}[h]                                  \vspace{-.5cm}
\centerline{\psfig{figure=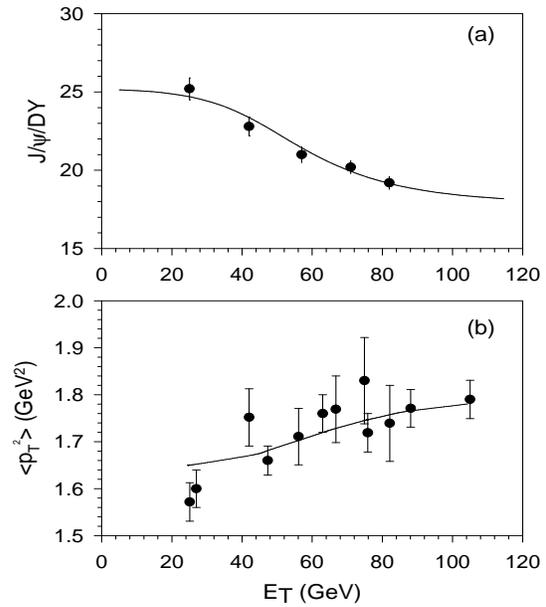,height=10cm,width=8cm}}
\vspace{-1.cm}  \caption{(a)  NA38  data  on  $E_T$ dependence of
$J/\psi$ suppression in S+U collisions. The solid line is  a  fit
obtained  in  the  'unconventional' nuclear absorption model. (b)
NA38 data on $p_T$ broadening in S+U along with the fit  obtained
in the 'unconventional' nuclear absorption model.} \end{figure}

$p_T$  broadening  of  $J/\psi$'s in AA collisions depends on two
parameters,  (i)  $<p^2_T>_{NN}$,  the  mean  squared  transverse
momentum  in  NN  collisions,  a measurable quantity and (ii) the
product of the gluon-nucleon cross section and the average parton
momentum broadening per collision,  $\sigma_{gN}\delta_0$.  Since
gluons  are  not  free,  the  second  quantity is essentially non
measurable. We obtain $\sigma_{gN}\delta_0$ from  a  fit  to  the
NA38  $p_T$  broadening data \cite{na38} in S+U collisions at 200
GeV/c. $<p^2_T>_{NN}$ at corresponding energy is known  from  NA3
experiment,  $<p^2_T>_{NN}  =  1.23  \pm  0.05$  \cite{na3}.  The
$E_T-b$ correlation parameters, $a$ and $q$  for  S+U  collisions
are,  $a$=3.2  and  $q$=0.74  GeV  \cite{vo99}.  To show that the
present 'unconventional' nuclear absorption model also reproduces
the centrality dependence of $J/\psi$ over Drell-Yan ratio in S+U
collisions, in Fig.1a, we have  compared  our  results  with  the
experimental   data.  We  have  neglected  the  effect  of  $E_T$
fluctuations in S+U collisions. The agreement  between  data  and
theory  is  good.  In Fig.1b, NA38 experimental data on the $E_T$
dependence of $p_T$ broadening are shown. The solid line is a fit
to the data,  obtained  with  $<p^2_T>_{NN}$=  1.23  (fixed)  and
$\sigma_{gN}\delta_0$    =    $0.442   \pm   0.056$.   Value   of
$\sigma_{gN}\delta_0$ agrees closely with the value  obtained  by
Kharzeev et al \cite{kh97} in the conventional nuclear absorption
model  and  also  with  the  value  obtained in the comover model
\cite{ar99}.

\begin{figure}[h]
\centerline{\psfig{figure=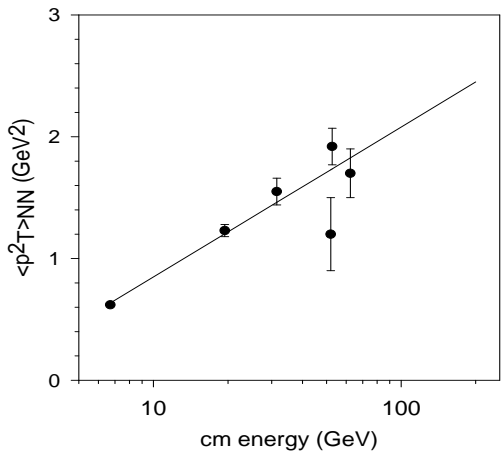,height=10cm,width=8cm}}
\vspace{-3cm}  \caption{Experimental $<p^2_T>_{NN}$ as a function
of cm energy along with the fit with Eq.\ref{6}.} \end{figure}

$<p^2_T>_{NN}$   increases   weakly   with   energy.   To  obtain
$<p^2_T>_{NN}$ for Pb+Pb collisions at 158 GeV/c, we have  fitted
the  existing  experimental  data \cite{na3,ba78,nag75,cl78} with
logarithmic energy dependence,

\begin{equation}  \label{6}  <p^2_T>_{NN}  =  a  + b \ln \sqrt{s}
\end{equation}

In Fig.2, experimental data along with the fitted curve is shown.
From the above parameterization, we obtain, $<p^2_T>_{NN}$
=1.15 $GeV^2$, for Pb+Pb collisions at CERN SPS. As we intend  to
predict  $p_T$  broadening at RHIC energy, $<p^2_T>_{NN}$ at RHIC
energy ($\sqrt{s}$=200 GeV)  is  also  obtained  from  the  above
parameterization.  At  RHIC energy, $<p^2_T>_{NN}$ =2.45 $GeV^2$.
However, we must warn our reader to treat the above  number  with
caution.  The  experimental  data  being  limited to 60 GeV only,
extrapolation to RHIC energy is unreliable.

\begin{figure}[h]
\centerline{\psfig{figure=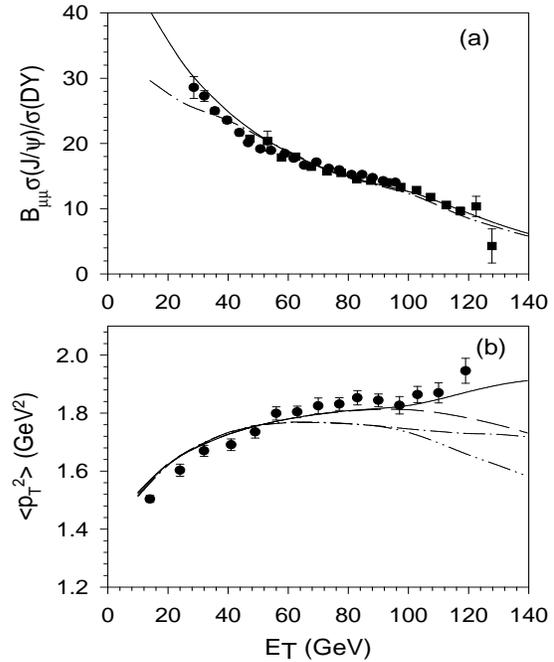,height=10cm,width=8cm}}
\vspace{-.5cm}  \caption{  (a) NA50 data on the centrality dependence
of the ratio, $J/\psi$ over Drell-Yan, is compared  with  the  
'unconventional' QCD
based  nuclear  absorption  model  (solid  line)  and  QGP  based
threshold model (dashed line). (b) NA50 data on $p_T$  broadening
of $J/\psi$'s in Pb+Pb collisions. The solid and dashed lines are
the  prediction in the  'unconventional' nuclear absorption model,
with and without the effect of $E_T$ fluctuations on $N_{AB}(b)$.
The   prediction   in   the    QGP    based    threshold    model
{\protect\cite{bl00}},  with  and  without the $E_T$ fluctuations
are shown as the dash-dot and dash-dot-dot lines.} \end{figure}

For  completeness purpose, in Fig.3a, we have shown the NA50 data
\cite{na50a} on the centrality dependence of the ratio,  $J/\psi$
over Drell-Yan. The solid line is the fit obtained to the data in
the 'unconventional' QCD based nuclear absorption model. The data
are  well  explained  in the model. In Fig.3a, we have also shown
the $J/\psi$ suppression obtained in the QGP base threshold model
\cite{bl00}.  In  the  threshold  model,  in  addition   to   the
conventional  nuclear  absorption,  an  anomalous  suppression is
included such that all the  $J/\psi$'s  are  suppressed  above  a
critical  density  $n_c$.  The  dashed  line  is  obtained in the
threshold model with $n_c$=3.7  $fm^{2}$.  \cite{bl00}.  It  also
gives satisfactory description to the data. Centrality dependence
of  $J/\psi$  suppression could not distinguish different natures
of absorption.

In  Fig.3b,  we  have compared the centrality dependence of $p_T$
broadening in the model with the NA50 experiment \cite{na50b}. We
have       used       $<p^2_T>_{NN}$=1.15       $GeV^2$       and
$\sigma_{gN}\delta_0$=0.442.  The  solid and dashed lines are the
$p_T$  broadening  with  and  without   the   effect   of   $E_T$
fluctuations   on   $N_{AB}(b)$.   When   the   effect  of  $E_T$
fluctuations is not taken  into  account,  the  $p_T$  broadening
continues  to  increase  with $E_T$ till 100 GeV (the knee of the
$E_T$ distribution). Thereafter, $p_T$ broadening decreases.  The
behavior  is  unlike  the  $p_T$ broadening in the 'conventional'
nuclear absorption model, rather more like the behavior in a  QGP
\cite{kh97}.  In  the  conventional  nuclear absorption model, at
large $E_T$,  $p_T$  broadening  saturates  while  in  a  QGP  it
decreases.  Indeed,  the different centrality dependence of $p_T$
broadening in nuclear and  in  QGP  medium  led  Kharzeev  et  al
\cite{kh97}  to  suggest  'decreasing  $p_T$  broadening at large
$E_T$' as a signal of QGP. We find that 'unconventional'  nuclear
absorption  model also produces a 'decreasing $p_T$ broadening at
large $E_T$'.
   At   large   $E_T$,  $J/\psi$'s  are  largely  suppressed  and
$<p_T^2>$ decreases. Decrease of $<p^2_T>$ at large $E_T$ can not
be considered as a signal of deconfinement phase transition.

Centrality dependence of $<p^2_T>$ at large $E_T$ is changed when
the   effect   of   $E_T$  fluctuations,  on  average  number  of
gluon-nucleon collisions, is taken into account.  The  decreasing
trend  of  $<p^2_T>$  beyond  100 GeV is changed to an increasing
trend (the solid line). Effect of $E_T$ fluctuations  essentially
increases  the  average  number  of gluon-nucleon collisions at a
fixed impact parameter and counter balance the large  suppression
effect  beyond the knee of the $E_T$ distribution. NA50 data also
shows an increasing trend beyond 100 GeV.  The  model  reproduces
the data very well (within 2\%).

We  have  also analysed the data in the QGP based threshold model
\cite{bl00}.  Huefner  et   al   \cite{hu02}   calculated   $p_T$
broadening  in  the threshold model (Fig.2 in ref\cite{hu02}) and
found that the NA50 data could not be fitted in the  model.  They
did not take into account the effect of $E_T$ fluctuations on the
average  number  of gluon-nucleon collisions, which we have seen,
is important for explaining the data. In Fig.3b, the dash-dot-dot
line is the $p_T$ broadening in the threshold model, without  the
effect  of  $E_T$ fluctuations on $N_{AB}$. As in \cite{hu02} the
model can not explain the data. However, at  low  $E_T$,  the  it
predict   $p_T$   broadening   in   close   agreement   with  the
'unconventional'  QCD  based  nuclear  absorption  model.  It  is
expected,  as  in  peripheral  collisions  (low $E_T$) QGP is not
produced and $J/\psi$'s are suppressed in  nuclear  medium  only.
When  the  effect  of  $E_T$ fluctuations on $N_{AB}(b)$ is taken
into account (the dash-dot  line),  even  though  the  difference
between  the  theory  and experiment is lessened, the model still
fails to explain the data. At large $E_T$ the model  predict  4\%
less  $p_T$  broadening,  also the upward trend beyond 100 GeV is
not reproduced. However, we note that the $p_T$ broadening in the
threshold model and in the  nuclear  absorption  model  are  very
close  to  each  other, the difference is less than within 3\% or
so. Since the NA50 data are very accurate, it can distinguish the
small difference between the two models. We may mention that NA50
$p_T$ broadening data were analysed by Armesto et al \cite{ar99},
in the comover model. The comover model also gives very good  fit
to  the  data,  only  the  increasing trend after $E_T$= 100 GeV,
could not be reproduced. They also neglected the effect of  $E_T$
fluctuations  on  the average number of gluon-nucleon collisions,
which give that tendency.

We  now  give prediction for $p_T$ broadening at RHIC energy. For
RHIC energy, parameters $a$ and $q$ in the $E_T - b$  correlation
are  taken  as;  a=1.97  and  q=0.46  GeV \cite{ch02a}. They were
obtained by fitting rescaled $E_T$  distribution  data  in  Pb+Pb
collisions  at  CERN  SPS.  At  RHIC  energy,  the so called hard
component,  which  is  proportional  to  the  number  of   binary
collisions,  appear.  Model  dependent calculations indicate that
the hard component grows from 22\% to 37\% as the energy  changes
from  56  GeV to 130 GeV \cite{kh01}. In our calculation, we have
used 37\% hard scattering component.  In  Fig.4a,  the  predicted
centrality  dependence  of  the  ratio of $J/\psi$ over Drell-Yan
ratio is shown. The QCD based nuclear absorption  model  and  the
QGP  based threshold model, both predict nearly the same $J/\psi$
suppression. Different nature suppression is not distinguished.

\begin{figure}[h]
\centerline{\psfig{figure=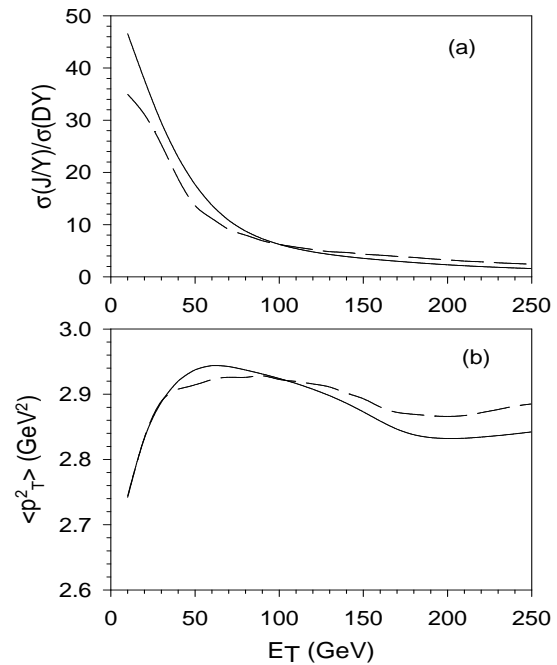,height=10cm,width=8cm}}
\vspace{-.5cm}  \caption{(a)  Predicted  centrality dependence of
$J/\psi$ over Drell-Yan ratio at RHIC. The solid and dashed lines
are the prediction in the QCD based nuclear absorption model  and
the QGP based threshold model. (b) Predicted centrality dependence
of $p_T$ broadening of $J/\psi$'s.} \end{figure}

In  Fig.4b,  the predicted $p_T$ broadening at RHIC are shown. We
have used $<p^2_T>_{NN}$
=2.45 ($GeV^2$) and $\sigma_{gN}\delta_0$=0.442. We again warn that
these  numbers  are  approximate  only,  The solid and the dashed
lines are the predicted
$p_T$   broadening   obtained  in  the  'unconventional'  nuclear
absorption model and in the QGP based threshold model, both  with
the  effect  of $E_T$ fluctuations on $N_{AB}$ included. Both the
models  give  similar  centrality  dependence   for   the   $p_T$
broadening.  After the initial rise with $E_T$, large suppression
forces the $p_T$ broadening to decrease, till 180 GeV(  the  knee
of the $E_T$ distribution). Beyond the knee, large suppression is
counter  balanced  by  the  effect  of  $E_T$ fluctuations on the
average number of gluon-nucleon collisions and  $p_T$  broadening
nearly  saturates.  The  predictions  in two models closely agree
(within  1.5\%)  with  each  other.  Given  the  uncertainty   in
$<p^2_T>_{NN}$  and  $\sigma_{gN}\delta_0$, such small difference
may not be distinguished experimentally. The results suggest that
at RHIC energy, centrality  dependence  of  $p_T$  broadening  of
$J/\psi$,  like the centrality dependence of $J/\psi$, may not be
able to distinguish a deconfinement phase transition.

To  summarize, we have analysed the NA50 data on $p_T$ broadening
of  $J/\psi$'s.  The  data  were  analysed  in  the  QCD   based,
'unconventional'  nuclear absorption model \cite{ch02} and in the
QGP based threshold model \cite{bl00}.  It  was  shown  that  the
$E_T$ fluctuations at a fixed impact parameter plays an important
role  in  explaining  the $p_T$ broadening data. If the effect is
not incorporated, both the  models  predict  a  decreasing  $p_T$
broadening  at  large $E_T$, contrary to the experiment. When the
effect is included, the 'unconventional' nuclear absorption model
could explains the data without any free parameter. The QGP based
threshold model could not explain the data. At  large  $E_T$,  it
produces  4\%  less  $p_T$  broadening, also the increasing trend
beyond 100 GeV is not  reproduced.  The  analysis  also  indicate
that,  the  'visibly decreasing $p_T$ broadening at large $E_T$',
can not be considered as  a  probe  of  the  deconfinement  phase
transition.  "Unconventional'  nuclear absorption also produces a
decreasing $p_T$ broadening at large $E_T$. We have also obtained
prediction for centrality dependence of $p_T$ broadening at  RHIC
energy.  At  RHIC,  both  the  models  predict very similar $p_T$
broadening. We conclude that $p_T$ broadening of  $J/\psi$'s  can
not probe the deconfinement transition at RHIC energy also.


\begin{references}  \bibitem[*]{byline}e-mail:akc@veccal.ernet.in
\bibitem{ma86} T. Matsui and H. Satz, Phys. Lett. B178,416(1986).
\bibitem{na50a}NA50 collaboration, M. C. Abreu {\em et al.} Phys.
Lett. B 477,28(2000). \bibitem{bl00} J. P. Blaizot,  P.  M.  Dinh
and    J.Y.   Ollitrault,   Phys.   Rev.   Lett.   85,4012(2000).
\bibitem{ch01} A. K. Chaudhuri, Phys.Rev. C64,054903(2001), Phys.
Lett. B527,80(2002) \bibitem{ca00} A. Capella, E. G. Ferreiro and
A. B. Kaidalov, hep-ph/0002300, Phys. Rev. Lett. 85,2080  (2000).
\bibitem{ch02}  A. K. Chaudhuri, Phys. Rev. Lett.88,232302(2002).
\bibitem{ch02a} A. K. Chaudhuri, nucl-th/0207082. \bibitem{na50b}
M. C. Abreau et al, Phys. Lett. B499,85 (2001). \bibitem{kh97} D.
Kharzeev, M.  Nardi  and  H.  Satz,  Phys.  Lett.  B405,14(1997).
\bibitem{qiu98}  J. Qiu, J. P. Vary and X. Zhang, hep-ph/9809442,
Nucl. Phys. A698, 571 (2002). \bibitem{be94} C. J. Benesh, J. Qiu
and J. P. Vary, Phys. Rev. C50, 1015 (1994). \bibitem{na38}  NA38
collaboration,   C.  Baglin  et  al,  Phys.Lett.B262,362  (1991).
\bibitem{na3}  NA3  collaboration,J.  Badier  et  al,  Z.   Phys.
C20,101(1983).  \bibitem{vo99}  R.  Vogt, Phys. Reports, 310, 197
(1999). \bibitem{ar99} N. Armesto, A. Capella and E. G. Ferreiro,
Phys. Rev. C59, 395 (1999).  \bibitem{ba78}A.  Bamberger  et  al,
Nucl.  Phys.  B134,1  (1978), \bibitem{nag75}E. Nagy et al, Phys.
Lett.  B60,96(1975).  \bibitem{cl78}A.  G.  Clark  et  al,   Nucl
Phys.B142,29(1978).  \bibitem{na38_98}NA38  Collaboration,  M. C.
Abreu et  al.  Phys.  Lett.  B423,  207  (1998)  \bibitem{hu02}J.
Huefner   and   P.  Zhuang,  nucl-th/0208004.  \bibitem{kh01}  D.
Kharzeev    and    M.    Nardi,    Phys.     Lett.B507,121(2001).
\end{references}
 \end{document}